\documentclass[sigconf,nonacm]{acmart} 
\usepackage{graphicx} % Required for inserting images
\usepackage{listings}

\usepackage[most]{tcolorbox}
\definecolor{pastelyellow}{RGB}{255,241,206}
\definecolor{pastelblue}{RGB}{231,234,241}
\definecolor{pastelgrey}{RGB}{230,230,230}

\lstdefinestyle{inputstyle}{
    language=SQL,
    frame=single,
    rulecolor=\color{blue},
basicstyle=\ttfamily\small
}

\lstdefinestyle{exastyle}{
    language=SQL,
    frame=single,
    rulecolor=\color{gray},
basicstyle=\ttfamily\small
}

\lstdefinestyle{rulestyle}{
    language=Java,
    frame=single,
    rulecolor=\color{teal},
basicstyle=\ttfamily\small
}

\usepackage{subfigure}

\usepackage{xargs}
\usepackage[pdftex,dvipsnames]{xcolor} 
\usepackage[colorinlistoftodos,prependcaption,textsize=tiny]{todonotes}
\newcommandx{\urgent}[2][1=]{\todo[linecolor=red,backgroundcolor=red!25,bordercolor=red,#1]{#2}}
\newcommandx{\prove}[2][1=]{\todo[linecolor=teal,backgroundcolor=teal!25,bordercolor=teal,#1]{#2}}
\newcommandx{\add}[2][1=]{\todo[linecolor=blue,backgroundcolor=blue!25,bordercolor=blue,#1]{#2}}
\newcommandx{\kontrollieren}[2][1=]{\todo[linecolor=green,backgroundcolor=green!25,bordercolor=green,#1]{#2}}
\newcommandx{\sophisticate}[2][1=]{\todo[linecolor=violet,backgroundcolor=violet!25,bordercolor=violet,#1]{#2}}

\title{NovaFilter – A Rule-Based ETL Staging Framework for
Fault-Tolerant Relational Databases}
\author{Meike Klettke}
\affiliation{%
  \institution{University of Regensburg}
  \city{Regensburg}
  \country{Germany}}
\email{meike.klettke@ur.de}
\date{May 2026}

\begin{document}

\begin{abstract}
NovaFilter is a completely new approach to deriving structured relational data from heterogeneous input data in different data formats. It combines consistency checks, rule-based data transformation, and a human-in-the-loop example discussion. Using these three components, NovaFilter stores all transformation knowledge in an explicit rule base, which is continuously expanded through the example discussion with the user. The resulting data transformation process is deterministic, transparent, and reusable.
As a result, the method can replace conventional ETL scripts and offers a completely new strategy for transforming data between different data models and schemas.

%NovaFilter is a completely new approach to derive structured relational data from heterogeneous input data in various formats. It combines rule-based constraint checking, rule-based data transformation, and a human-in-the-loop example discussion. With these three components, 
%NovaFilter externalizes all transformation knowledge into an explicit rule base that continuously grows from user-confirmed example discussions. The resulting data transformation process is deterministic, explainable and reusable.
%The method can replace conventional ETL scripts and suggests a completely new strategy for transforming data between different data models and schemas.

Its main advantage is that users without programming skills (e.g. domain experts) can answer the example discussion and thus it is a building block for democratization of ETL script development. 
\end{abstract}

\maketitle

\section{Motivation}

\begin{figure}[h]
    \centering
\includegraphics[width=0.61\columnwidth]{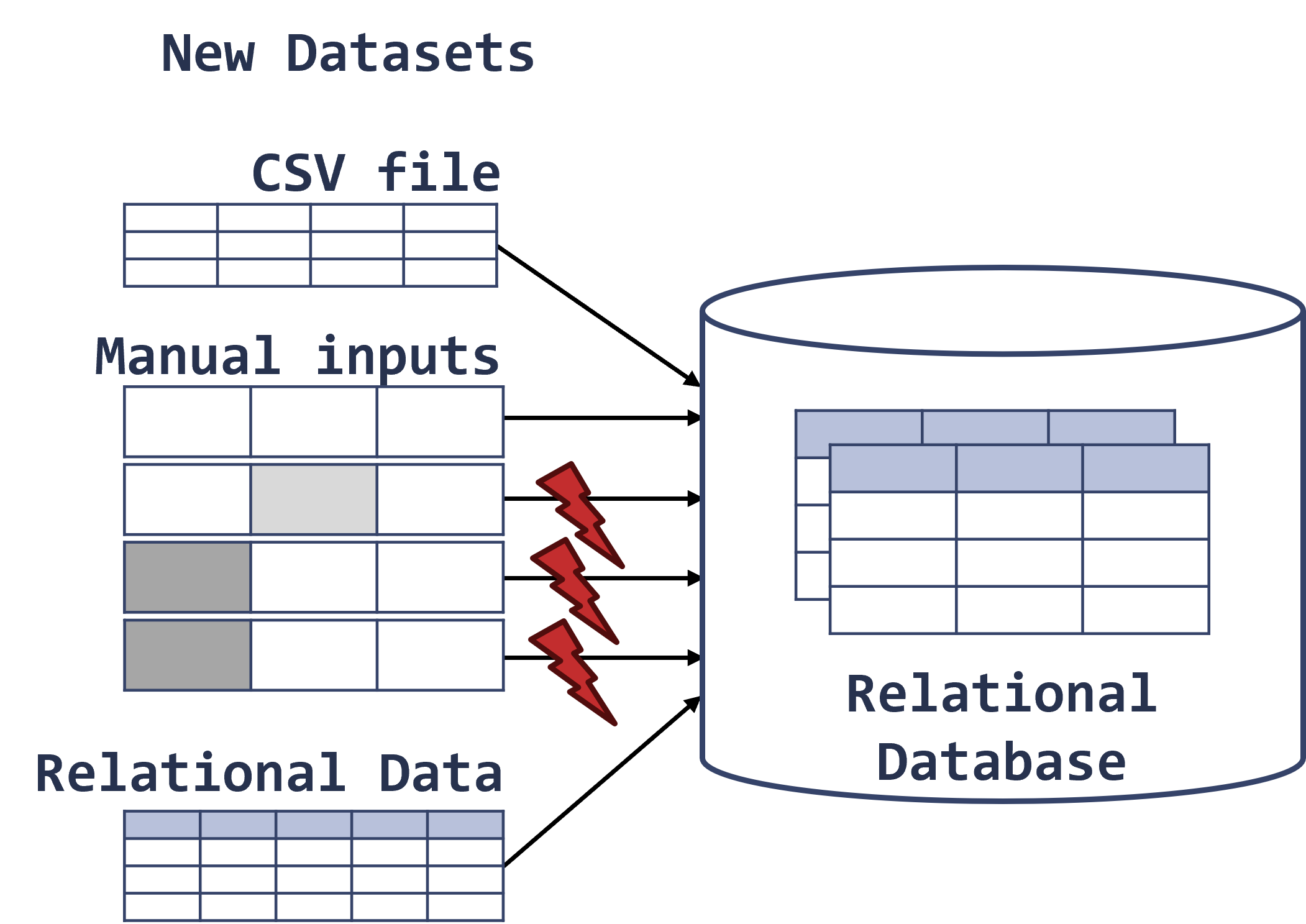}
\caption{Inserting noisy or incomplete data into a relational database}
\label{fig:problem}
\end{figure}

Relational databases are the backbone of enterprise information systems  for several decades. Their main strengths are a clearly defined schema, strict data types, and guaranteed integrity constraints. These mechanisms ensure high data quality but pose challenges when data are imported from \textit{heterogeneous sources}.
In practical scenarios, data are obtained from manual data entry, web forms, external files, or external systems. The difference between input data characteristics and the structure of the target relational database can cause errors during data ingestion (see Figure \ref{fig:problem}). Typical problems when loading such data into a relational database include:

\begin{itemize}
\item Different \textit{data types} (like strings instead of numeric values),

\item Violated \textit{data type constraints} (like violated 
range constraints or 
string lengths),
or enumeration values), 

\item \textit{Missing mandatory attributes},

\item \textit{Additional attributes}, 

\item Different \textit{attribute names} (like misspelling, abbreviations, or different languages in attribute names), 

\item Inconsistent or missing \textit{references}, and 
\item \textit{Combinations} of these 
differences. 
\end{itemize}

\begin{figure}[ht]
\includegraphics[width=0.95\columnwidth]{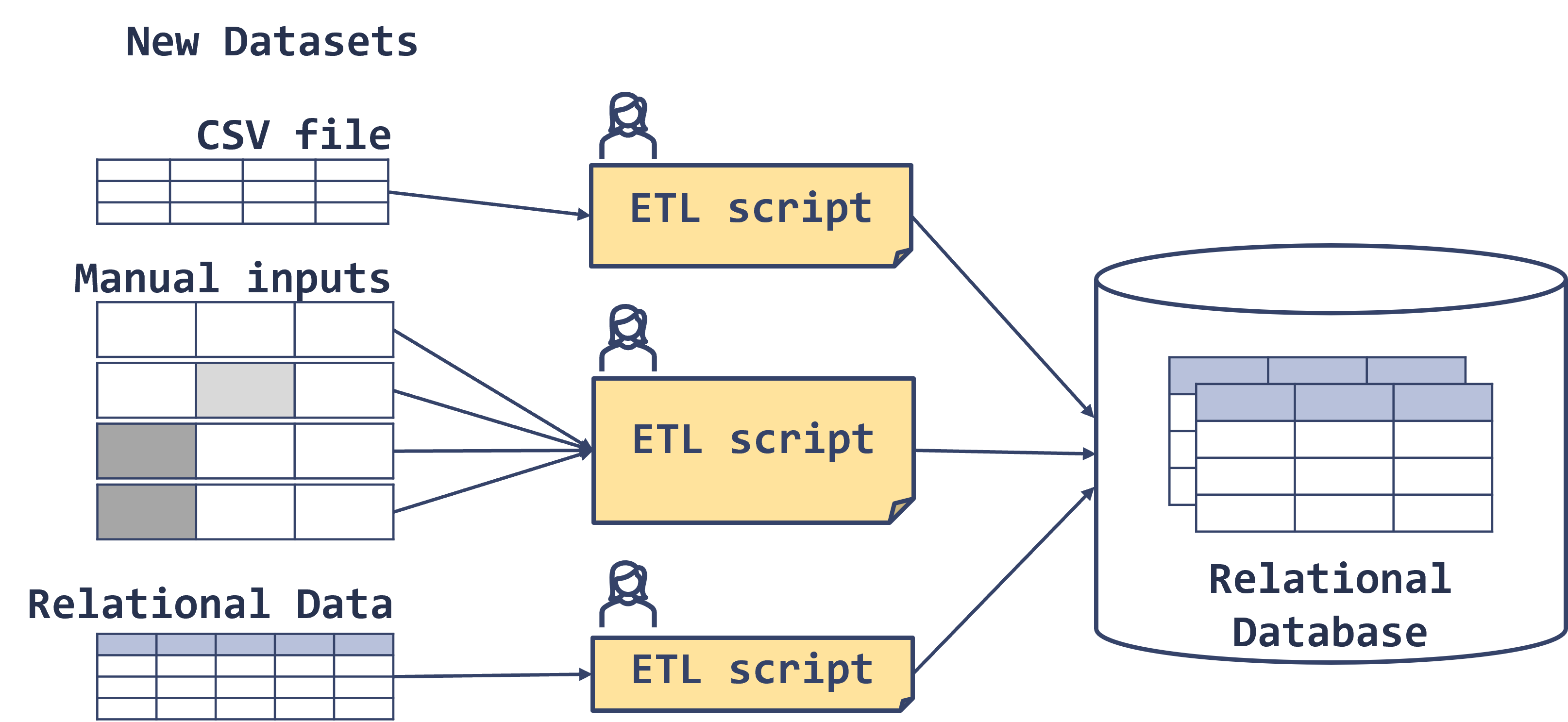}
\caption{ETL scripts for data 
%error correction and 
transformation}
\label{fig:etl}
\end{figure}

Conventionally, \textit{ETL scripts} are implemented and used to solve this data ingestion task (see Figure \ref{fig:etl}).
Their  development 
requires a large manual effort; 
they are time-consuming 
and costly to implement and maintain.
Another challenge is that ETL scripts are often incomplete, since it is nearly impossible to \textit{predict all possible data variants}, especially when loading heterogeneous data or entering data manually. 

%In contrast, \textit{JSON database systems} have been developed to store data in different variants. They allow schemaless or partially schema-controlled storage and thus enable \textit{lossless ingestion of heterogeneous data}. However, without schema control, JSON database   are not suitable as backend of critical applications, particularly because the data heterogeneity makes NoSQL queries more complex and the results less reliable.

Currently, numerous research studies 
%(\cite{}) 
are exploring how GenAI, especially LLMs can be integrated into the data transformation workflow. Although this often leads to excellent results, a process automated in this way carries the risk of erroneous translations, and we believe that it cannot be applied to a task as important as data transformation without a validation loop.
The well-known reason for this lies in the “hallucinations” of these systems; furthermore, missing domain expertise can lead to incorrect mappings during the transformation. We view the lack of explainability of the results as an inherent problem with the method.
Furthermore, using LLMs carries the same \textit{risk as outsourcing} in general: the results produced by LLMs can change over time (e.g., with new LLM releases), and the costs of the transformation have to be paid in each usage and these costs can also change over time.
 
\begin{figure*}
    \centering
\includegraphics[width=0.75\linewidth]{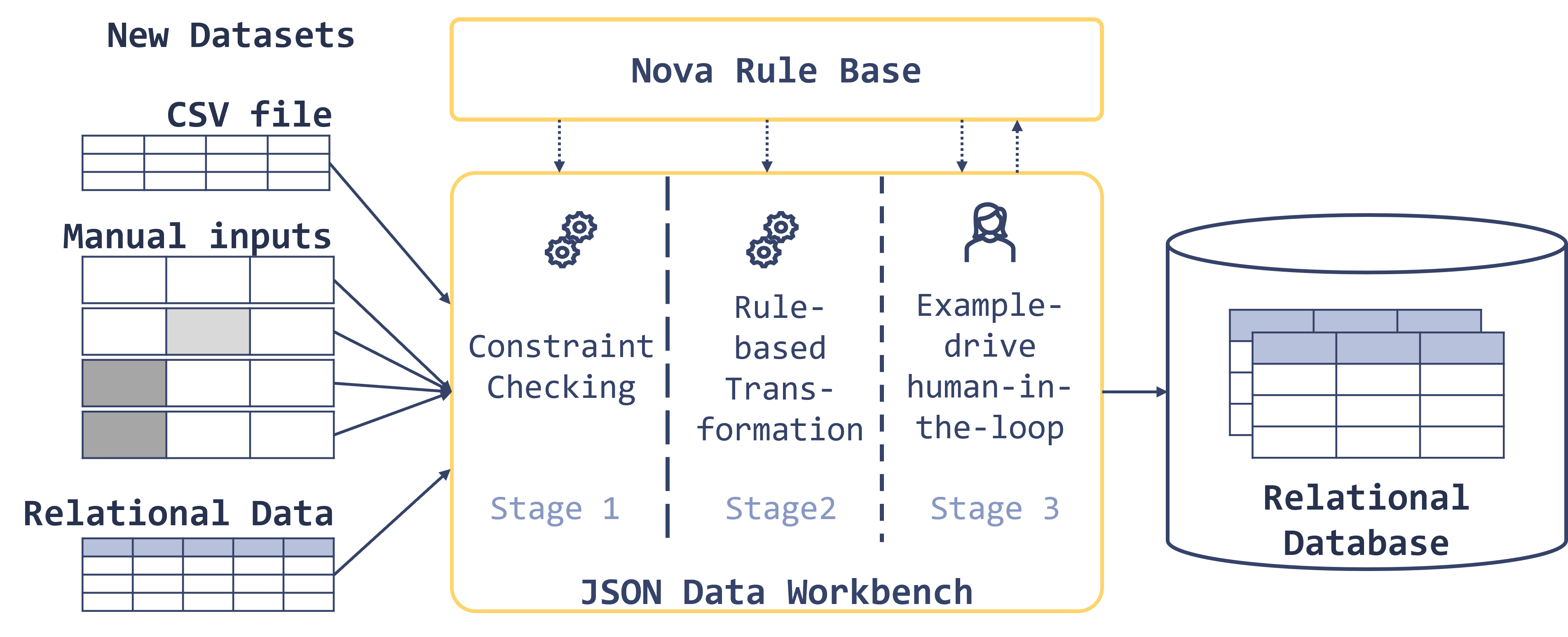}
\caption{Transformation Process in  NovaFilter 
%- Consistency Checking and Transformation Tool for Relational Databases 
\label{fig:novaprocess}}
\end{figure*}

We therefore propose a completely different approach, which combines rule-based methods and human-in-the-loop example discussion. the three steps of the transformation process are shown in Figure \ref{fig:novaprocess}.  Our approach uses a \textit{semistructured staging area} in which JSON documents store all incoming data and enable a \textit{lossless ingestion of heterogeneous input data}. Before loading all datasets into the relational database, the datasets are stepwise validated and transformed (see Figure \ref{fig:novaprocess}, Step 1 and 2. Instead of ETL approaches or LLM generated scripts, we develop a completely deterministic, rule-based method with \textit{constraint checking} and \textit{rule-based transformation} for the ETL tasks. 

 This ensures complete transparency of the data transformation performed. Only in cases where an automated process does not work (e.g., where there are no applicable transformation rules) is a human-in-the-loop example discussion conducted (see Figure \ref{fig:novaprocess}, Step 3).
The aim of this project is to combine the best of both worlds:
\begin{itemize}
    \item  \textit{Flexibility} of a \textit{semi-structured staging area} (using JSON documents) and 
\item  \textit{Reliability and constraints guarantees} of \textit{relational databases}.
\end{itemize}

\noindent
With our approach, we will make the following contributions: 
\begin{enumerate}
\item Automatic data translation  between different data models. 
\item Rule-based constraints check and data transformation.
\item Example discussion with users to derive new transformation rules.
\item Design, continuous adaption, and usage of a rule base to guarantee that all patterns are only discussed once with a user. 
\end{enumerate}

The rest of the article is organized as follows: 
In Section \ref{sec:architecture}, we introduce the system architecture, in Section \ref{sec:Process} the overall process is demonstrated with an example, and in Section \ref{sec:Implementation} the implementation is described. In the following Section \ref{sec:RelatedWork}, we give an overview on the State-of-the-Art which we can use for the approach. 

%\noindent
%We develop a novel \textit{rule-based approach}  to insert heterogeneous data into relational databases (see Figure \ref{fig:architecture}). 
%The core idea is a JSON Data Workbench to load all new datasets into the \textit{staging area} in a \textit{flexible JSON format}.  

%The JSON schema of the data in the workbench as well as the \textit{validation rules} are be derived from the \textit{relational target database} and stored in the Nova Rule Base. 
%In the workbench, the datasets are \textit{validated} with the available rules (see Figure \ref{fig:architecture}, Stage 1) and, if necessary, \textit{transformed} using the available rules (Stage 2). Only if no transformation rules are applicable (for example, when new data sets are to be loaded), a \textit{human-in-the-loop approach} is started (in Stage 3). In it, the transformation of a sample dataset is discussed with the user, and a new transformation rule is derived and added to the Nova Rule Base. 

\section{System Architecture of NovaFilter \label{sec:architecture}}

\begin{figure*}
    \centering
\includegraphics[width=0.85\linewidth]{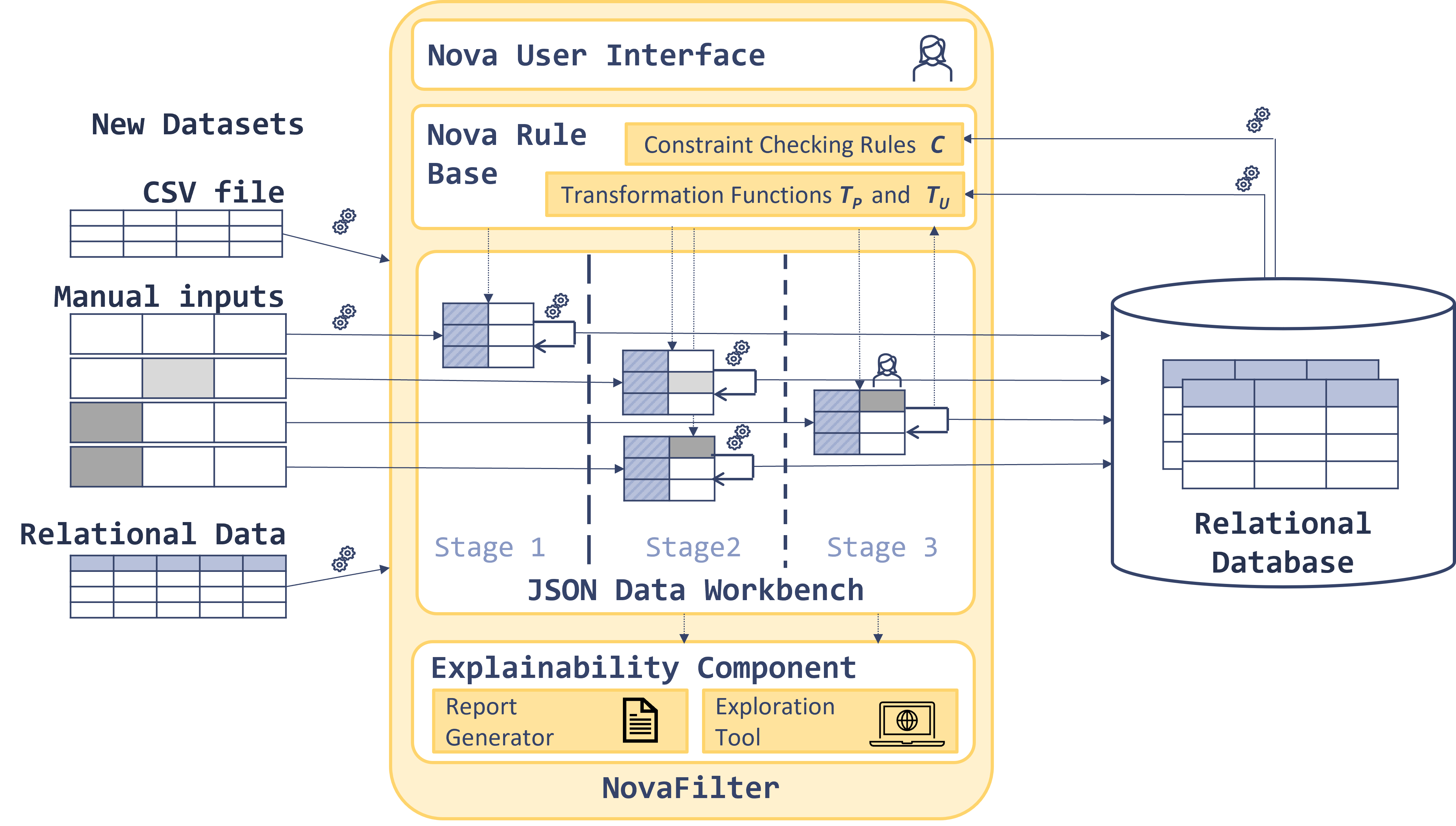}
\caption{System Architecture of NovaFilter 
%- Consistency Checking and Transformation Tool for Relational Databases 
\label{fig:architecture}}
\end{figure*}

NovaFilter is a generic, application-in\-de\-pen\-dent framework that acts as a pre-filter for relational insert and update operations. It is entirely rule-based. 
%In contrast to data transformation by using LLMs \cite{} approaches which is currently often applied, 
We only rely on \textit{predefined rules}, extracted rules from the target schema and \textit{user-defined rules} which are inferred from a sample discussion with users. 
The architecture of NovaFilter is shown in Figure \ref{fig:architecture} and includes the transformation process which we have already seen in Figure \ref{fig:novaprocess}. The architecture consists of the following main components.  
%In the following, we describe all components of the approach. 
%We start in the middle, with the Staging Area, followed by all other parts:

\paragraph{Nova Rule Base:} The rule base of the system provides \textit{for each target database} the 
\begin{itemize}
\item set of rules for consistency checks $C$,
\item the predefined transformation functions $T_P$ and \item the user-defined transformation functions $T_U$. 
\end{itemize}
These three rule sets serve as knowledge base for the entire transformation process.

\paragraph{JSON-Based Staging Area:} 
The Staging area is the central component of the approach. Here, all new datasets are loaded as JSON documents. Before the documents can be inserted into the target database, they pass three different stages: 

\begin{itemize}
\item \textit{Stage 1 --Constraint Checking:}
Structural and domain constraints (like property names, data types and value ranges) are extracted %automatically 
from the target relational schema and are stored in the Nova Rule Base. 
For each new dataset, all relevant  consistency rules $C$ are tested. If one rule is not fulfilled, the process is continued in Stage 2.

\item 
\noindent
\textit{Stage 2 -- Data Transformation Process:} Here, a stepwise data transformation is realized. It uses all  transformation functions which are available in the Nova Rule Base,  
the \textit{predefined transformation functions in $T_P$}, which are automatically generated for each new target database and the \textit{user-defined transformation functions $T_U$}, which are derived from previously performed user interactions.

\item
\noindent
\textit{Stage 3 -- Human-in-the-Loop Approach:}
Data that cannot be corrected automatically are presented to a user, e.g. a domain expert. From their feedback \textit{new transformation functions} are created and added to $T_U$. 
Over time, the system thus incrementally improves its capabilities.
The \textit{Once-Only Principle}\footnote{The Once-Only Principle is inspired by the state governance of Estonia that each information is only inquired once from the citizen.} that we use requires that a new transformation rule is derived from each example. 

\end{itemize}

%In this staging area, the consistency check and the transformation process shall be realized. It uses the rules set of the approach (see next item).  

\paragraph{Explainability Component}: The application of all data transformation functions from $t \in T_P \cup T_U$ is logged and can either generate a report as log-file or be used for an interactive exploration tool, organized like a dashboard. 

\paragraph{Nova Monitoring Component}: This part of the tool controls the entire process, the selection of the target database and the input datasets and starts the transformation process.

\noindent
%Combining all parts of the architecture, the research questions \textbf{R1} (feasibility) \textbf{R2} (how fast can the system learn) and \textbf{R3} (computational overhead) can be answered. 

%Next, the data transformation process in NovaFilter is defined. 

\section{Transformation Process \label{sec:Process}}

In the following, we show the data transformation process in detail and explain each step in Stage 1, 2 and 3 (see Figure \ref{fig:architecture}) with our running example. Our sample requires several different transformation tasks (transformation between different data models and transformation between different schemas).  

\subsection{Input Data and Target Database Structure}
The \textit{target database} is a \textit{relational database}, its schema definition is needed for the transformation process. 
As running example, we are using the following schema definition of a PostgreSQL database. 

\begin{tcolorbox}[
title=Schema of the Target Database,
colback=pastelblue,
colframe=blue!30!black,
arc=2mm]
CREATE TABLE products (\\
  \hspace*{0.25cm} product\_id INT NOT NULL PRIMARY KEY,\\
  \hspace*{0.25cm}
  product\_name VARCHAR(150) NOT NULL,\\
  \hspace*{0.25cm}
  price NUMERIC(10,2) NOT NULL,
  \\
  \hspace*{0.25cm}
  stock\_quantity INT NOT NULL DEFAULT 0,\\
  \hspace*{0.25cm}
  is\_active BOOLEAN NOT NULL DEFAULT TRUE,
  \\
  \hspace*{0.25cm}
  created\_at TIMESTAMP NOT NULL \\
  \hspace*{0.5cm}DEFAULT  
    CURRENT\_TIMESTAMP,\\
  \hspace*{0.25cm}
  CONSTRAINT chk\_product\_price
    CHECK (price > 0) );
\end{tcolorbox}

Input data can have different data formats like \textit{relational data, graph data, JSON} or \textit{csv files}. In the current version, we can consider data from PostgreSQL (relational data model), MongoDB (JSON documents), and Neo4j (graphs), csv files are also possible
For our example, we use a \textit{JSON input file} with similar semantics and an obviously different structure.

\begin{tcolorbox}[
title=Input Data (JSON),
colback=pastelblue,
colframe=blue!30!black,
arc=2mm]
\{ \hspace{0.1cm}"product\_id": "1001", \\
  \hspace*{0.25cm}
  "product\_name": "Mechanical Keyboard", \\
  \hspace*{0.25cm}
  "category": "Electronics", \\
  \hspace*{0.25cm}
    "pricing": \{ \\
  \hspace*{0.5cm}
      "currency": "EUR",
      \\
  \hspace*{0.5cm}
      "amount": 89.99 \},
      \\
  \hspace*{0.25cm}
    "inventory": 120 \}
\end{tcolorbox}

In the staging area all input data are transformed into JSON documents. In the ingestion process, a \textit{flattening of the input data} takes place. This decision was made because the target structure is a relational database cannot store nested objects. 
In case of nested properties, we are only storing the leaf nodes and are replacing the property names through the complete path information, in the sample file \texttt{"currency"} through \texttt{"pricing/currency"} and 
\texttt{"amount"} through \texttt{"pricing/amount"}. This  automatically guarantees uniqueness of property names that well-formed JSON documents have to fulfill. 

 Using JSON offers the advantage that all input information can be stored, independently from their concrete structure and heterogeneity. 
 All important information from the source data files is preserved in the JSON documents. 
 In the next steps, this JSON document is stepwise validated and transformed.

\subsection{NovaFilter, Stage 1}

For each target database (relational database), consistency constraints $C$  and predefined transformation rules $T_P$ are generated.

The consistency constraints $C$ contain structural information (property name, required properties, datatype, value constraints like \texttt{min, max, regex} and \texttt{uniqueness} constraints) of the relational database. For our running example an excerpt looks like this:  

\begin{tcolorbox}[
title=Nova Rule Base -- Constraints,
colback=pastelyellow,
colframe=blue!30!black,
arc=2mm]
"products": \{  \\
  \hspace*{0.25cm}
  "properties": \{  \\
  \hspace*{0.5cm}
    "product\_id": \{
      "type": "int",  
      "default": "None"\}, \\
  \hspace*{0.5cm}
    "product\_name": \{  
      "type": "string", 
      "default": "None"\},
       \\
  \hspace*{0.5cm}
    ... \},\\
  \hspace*{0.5cm}
  "unique": ["product\_id"], \\
  \hspace*{0.5cm}
  "required": ["product\_id", "product\_name", 
               "price",\\
  \hspace*{1.5cm} "stock\_quantity"] \}
\end{tcolorbox}

Applying these to the input dataset, NovaFilter finds out that 
the property \texttt{product\_id}  is available and the datatype does not match the definition, whereas \texttt{product\_name} is available and the datatype match the definition.
The other required attributes \texttt{price, stock\_quantity, is\_active, created\_at} are missing. That's why the approach is continuing on Stage 2 with automatic transformations. 

\subsection{NovaFilter, Stage 2}

On this stage, the predefined functions $T_P$ and the user-defined functions from the previous steps $T_U$ are applied. 

Typecast operations belong to the predefined transformation rules $T_P$.
They are automatically provided for all target attributes.  

In our running example, we need a typecast transformation 
functions between \texttt{string} and \texttt{int} as the following rule base excerpt shows and use it to transform the value in the running example. Storing all values as strings is a frequent pattern in JSON documents, so we assume that the typecast function is often needed for JSON input files.

\begin{tcolorbox}[
title=Nova Rule Base -- Predefined Transformations,
colback=pastelyellow,
colframe=blue!30!black,
arc=2mm]
typecast("products", "product\_id", 
  "string", "int", \\ \hspace*{1.28cm}string2int())
\end{tcolorbox}

With this typecast operation, the value of \texttt{product\_id} that is currently a \texttt{string} is transformed into an \texttt{int} value.

Further transformations are mappings that can take place if the transformation rules are already available. 
For the first dataset,  this is not possible because the list of user-defined transformation rules $T_U$ is initialized as an empty set. 
Our running example cannot be loaded into the target database (at least the \texttt{price} is missing, that's why a human-in-the-loop process (Stage 3) is started. 

The last subtasks on this stage are checking of semantic constraints (\texttt{keys, foreign keys}). Both require a \textit{lookup in the relational database}. It is carried out in the staging area (before the data is loaded into the relational database). For a primary key, 
%the check 
it is 
%verifies 
verified that 
%whether 
the value not exists in the relational database. For a foreign key, 
%it verifies whether 
we prove that
the corresponding primary key exists. 
%This relatively costly operation is made available as an optional operation in Novafilter.

\subsection{NovaFilter, Stage 3} 

In this stage, all transformations that cannot be automated are discussed with a user. In particular, a user has to identify which values (from the input data set) should be assigned to the required attributes and have not yet been matched. 
In our case, these are \texttt{"price", 
 "stock\_quantity", "is\_active"}, and  \texttt{"created\_at"}.
If a user associates the value 89.99 as \texttt{price} and 120 as \texttt{stock\_quantity}, we can derive the following rules and add them to $T_U$. 

\begin{tcolorbox}[
title=Nova Rule Base -- User-defined Transformations,
colback=pastelyellow,
colframe=blue!30!black,
arc=2mm]
\{ "products": \{ \\ \hspace*{0.25cm}
    "aliases": \{ \\ \hspace*{0.5cm}
      "price": ["pricing/amount"], \\ \hspace*{0.5cm}
      "stock\_quantity":["inventory"] \}\\ \hspace*{0.25cm}
    ...  \} \}
\end{tcolorbox}

This mapping consists of a structure containing the \textit{name} of the \textit{relational target table}, the alias  information in the Nova Rule Base with 
keys representing the 
\textit{attribute names} of the relational target database and 
as values an array 
with the corresponding \textit{property names}  or \textit{path expressions}.

When discussing the next example with the user (if there is more than one dataset in the staging area, Level 3),  we proceed in the \textit{opposite manner} to that used in \textit{reinforcement learning}. We start with the datasets in which  \textit{only one constraint is violated}, so that the derivation of the new user-defined transformation rule is \textit{unambiguous}.
We can then apply these new user-defined transformation rules in $T_U$ to the next data sets, thereby reducing the number of cases in a data set that need to be discussed with the user.
 
The \textit{reversal of the reinforcement learning process} seems particularly well-suited for this purpose, as it does not minimize the number of cases to be discussed but guarantees the unambiguity of the derived transformation rules.   

Applying these rules, we generate the following JSON document: 

\begin{tcolorbox}[
title=Nova Data Stage 3,
colback=pastelgrey,
colframe=blue!30!black,
arc=2mm]
  \{ \hspace*{0.1cm} "product\_id": 1001,\\ \hspace*{0.25cm}
    "product\_name": "Mechanical Keyboard",\\ \hspace*{0.25cm}
    "category": "Electronics",\\ \hspace*{0.25cm}
    "pricing/currency": "EUR",\\ \hspace*{0.25cm}
    "price": 89.99,\\ \hspace*{0.25cm}
    "stock\_quantity": 120 \}
\end{tcolorbox}

This JSON document from the staging area fulfills all constraints of the target database and can now be \textit{loaded into the relational database}. Unlike the target relational database, the other attributes (\texttt{“is\_active”, “created\_at”}) are not available in the staging area (and were not included in the input data either), but default values are defined for these attributes in the schema definition of the target relational database. Therefore, the dataset can be loaded from the staging area into the relational database.

In the JSON document, there are some additional properties (\texttt{category, pricing\_currency}) which cannot be stored in the relational target database. In our sample case, these values are ignored, in general it is also possible that values from a JSON document are loaded into different target databases, e.g. whether the input data also contain additional information like review or more detailed technical product information etc.

\subsection{Transforming Further Examples}

The same method can be applied for the next datasets, let's take as an example the following.

\begin{tcolorbox}[
title=Input Data (JSON),
colback=pastelblue,
colframe=blue!30!black,
arc=2mm]
\{ "product\_id": "1004",\\
  \hspace*{0.25cm}
    "product\_name": "Docking Station",\\
  \hspace*{0.25cm}
    "category": "Computer",\\
  \hspace*{0.25cm}
    "retailprice": 219, \\
  \hspace*{0.25cm}
    "inventory": 15 \}
\end{tcolorbox}

In this input dataset, for the target  attribute \texttt{stock\_quantity} there is already an applicable transformation function in $T_U$. It is applied to transform the attribute \texttt{inventory} to \texttt{stock\_quantity}.  
But there is no applicable mapping function for \texttt{price} in $T_U$ that's why the discussion with the user is done for this attribute only and the rule base is added in the following. 

\begin{tcolorbox}[
title=Nova Rule Base -- User-defined Transformations,
colback=pastelyellow,
colframe=blue!30!black,
arc=2mm]
\{ "products": \{ \\ \hspace*{0.25cm}
    "aliases": \{ \\ \hspace*{0.5cm}
      "price": ["pricing/amount", "retailprice"], \\ \hspace*{0.5cm}
      "stock\_quantity":["inventory"] \}\\ \hspace*{0.25cm}
    ...  \} \}
\end{tcolorbox}

So, we see that each human-in-the-loop interaction is only executed once and extends the user-defined transformation rules which are stored in the Nova Rulebase. This means that, for this alias discussion, the maximum number of discussion steps corresponds to the number of different attribute name variants. If input data are relational tables, then we have at most $n$ discussion steps (where $n$ is the total number of attributes in the input data). 

\section{Implementation and Discussion \label{sec:Implementation}}

The implementation is done in Python, the current 
target databases is  PostgreSQL. 
Input data can be relational data (PostgreSQL), graph data (Neo4j), JSON (Mongodb), or csv files. 
Currently, Stage 3 is not implemented yet, the concrete user interface and dialogue construction is still open. 

%\textit{Combining all the pieces together} is, of course, a challenge in itself; ultimately, but we expect a real added value, enabling a suggested novel approach to ETL processes.   

%\subsubsection*{Why not using LLMs instead} LLMs are currently being tested for various database tasks and often deliver fascinating results in prototypes.Nevertheless, we do not plan to use LLMs, as their results are not yet explainable,reliable, or reproducible. 
%Furthermore, LLMs have the advantages and disadvantages of outsourcing in general: One relies on an external component and is dependent on its stability.

%Instead, we plan to develop a rule-based approach in WP 1 - WP 4, which also carries the risk that an incorrect transformation rule is extracted from the user interaction, but which documents all rules and thus explains the transformation process. The second loop (explanation component) is planned in WP 5 for this purpose.

%But in the architecture, there is a docking point where this technology could be integrated in a future step. Instead of the dialogue with the user to discuss examples (\textbf{WP 3}) an automated process could be used. That means we would replace the rules $T_U$ through $T_M$ (LLM suggested transformations).
%In this case, the planned explanation component from \textbf{WP 5} offers the possibility to see the effects of the rules $T_M$  and thus can monitor the transformation correctness.

\begin{figure}[ht]
\centering
\includegraphics[width=0.65\columnwidth]{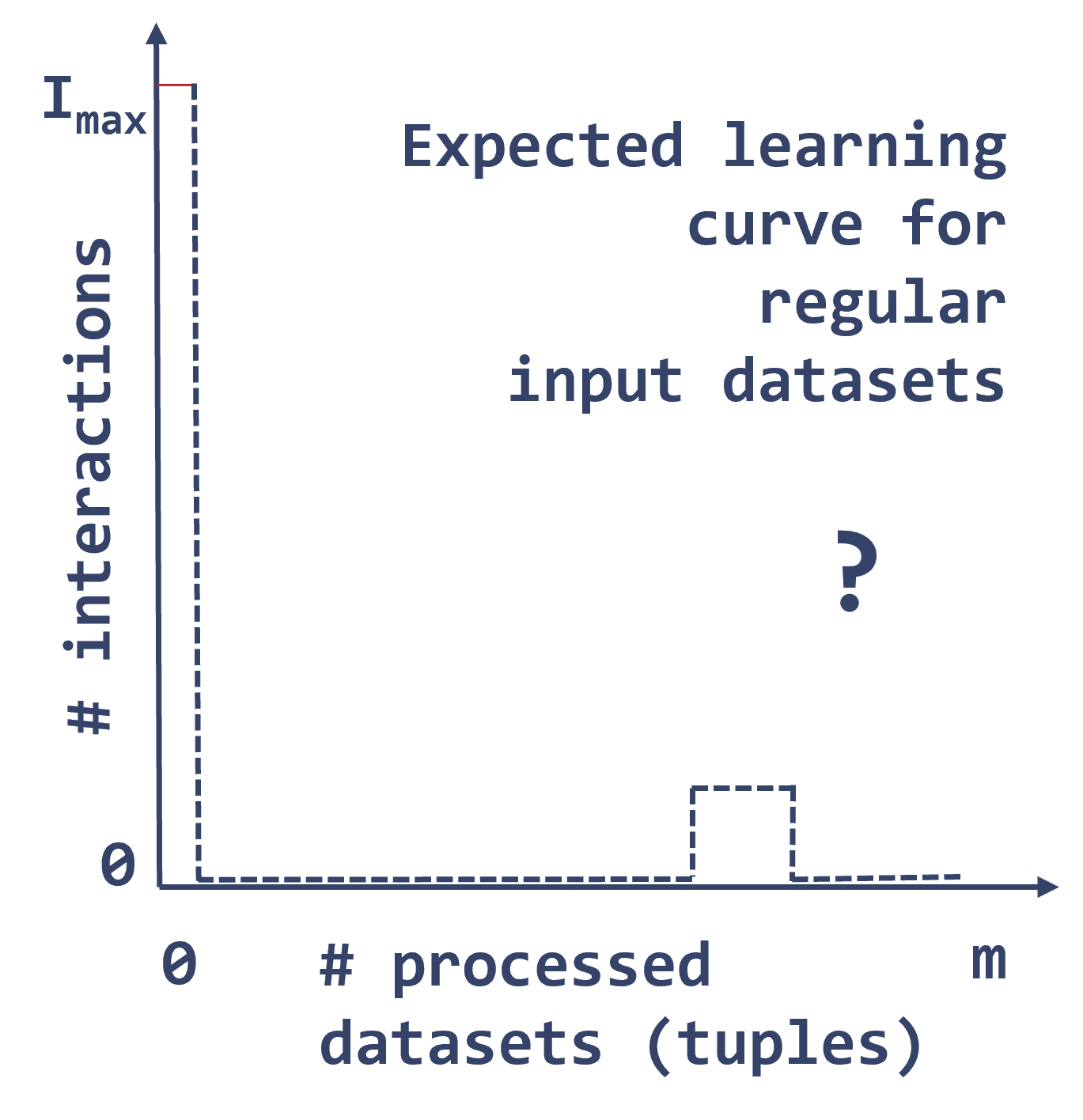}
\caption{Expected number of user interactions over time for regular data (relational input data)}
\label{fig:results_rel}
\end{figure}

\begin{figure}[ht]
\centering
\includegraphics[width=0.65\columnwidth]{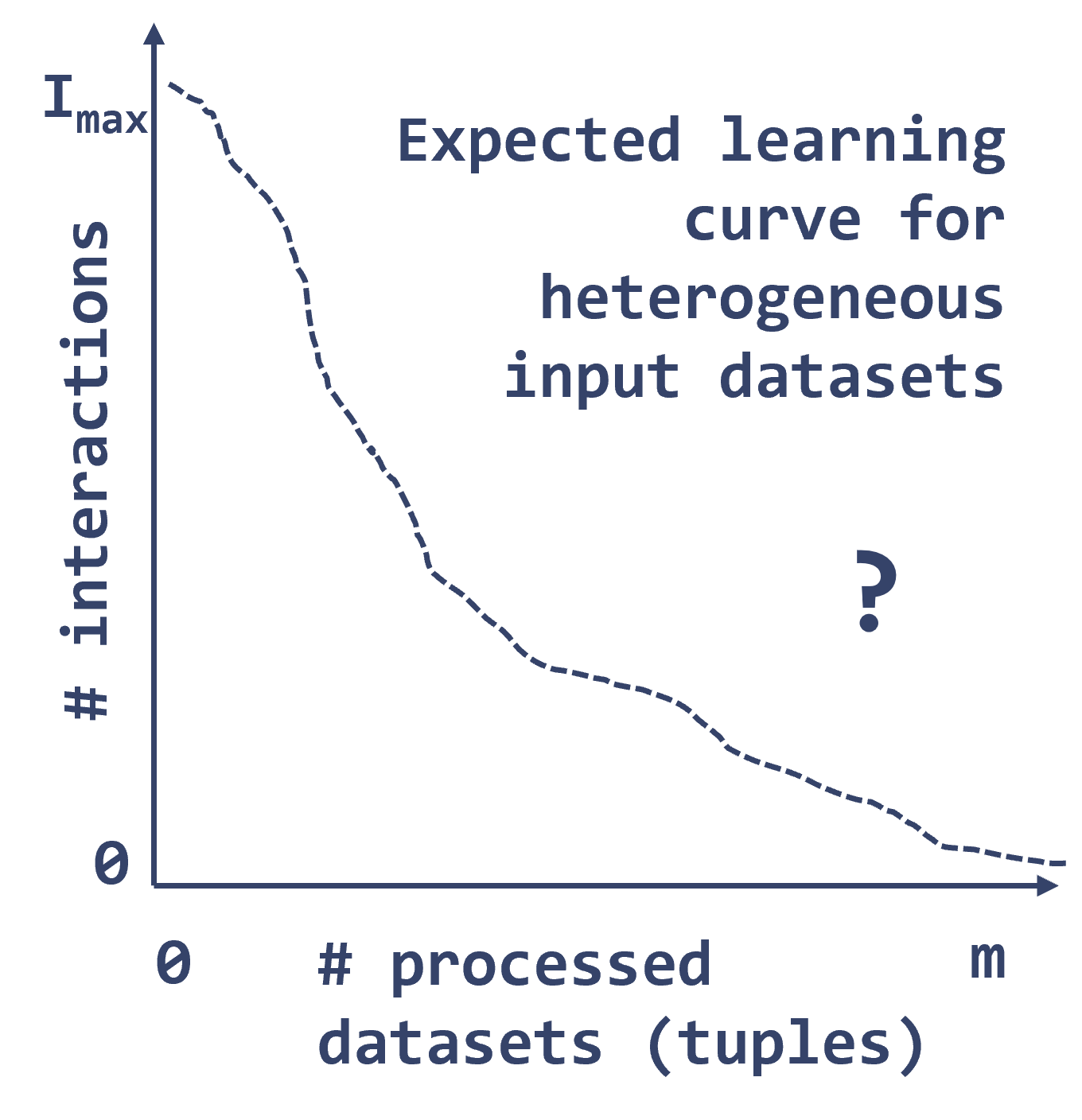}
\caption{Expected number of user interactions over time for heterogeneous input data (non-relational data like JSON documents or graph data)}
\label{fig:resultshet}
\end{figure}

We have not yet conducted comprehensive tests using input data with different characteristics, but we expect the following
number of interactions over time, which reflects the \textit{learning rate} (Figure \ref{fig:results_rel} and Figure \ref{fig:resultshet}).

%The prototype evaluation  shall test the interplay of all components and determine how many user interactions are needed to derive new transformation rules and adds it to $T_U$. 
%We want to find out whether there exists a \textit{cold-start problem} and how the dataset characteristics influence the number of  user-interaction steps. 

%We hope that we can often achieve for regular input data pattern as we see in Figure \ref{fig:results_relation}. 

%In this work package, we will evaluate the approach using concrete databases by correlating the number of data sets already transformed with the number of required user interactions. 

 %It defines how quickly the user-defined transformation rules  can be learned. 
For regular input data, which differs from the relational target database in some points, we expect that all data have the same structure and their transformation can be discussed with the user based on the first loaded tuple (see Figure \ref{fig:results_rel}). Only if another tuple exhibits different patterns (for example, contains a null value that was not present in the preceding tuples) may additional interactive steps be necessary in the transformation process.

For semi-structured data, we expect a rate as shown in Figure \ref{fig:resultshet} b). Here, different data variants are distributed in the input data and have to be discussed sequentially.
And, of course, we generally expect that the effort required for loading and transforming non-relational data will be higher; this is simply due to the nature of the data—its heterogeneity.

%For the validation, we will among available data sets also use the tool GouDa \cite{DBLP:conf/deem/RestatBCS22} for  generating data for different scenarios.

%With this validation, we want to answer our research question  \textbf{R2} (learning curve). For answering \textbf{R3} (performance overhead), especially through WP2, Stage 1, we plan detailed runtime measurements for different input data models, data characteristics and different database sizes.  

\section{Related Work \label{sec:RelatedWork}}

Since data transformation and data integration parts are evergreens in the field of data engineering, there are many existing approaches on which we build. 

\paragraph{ETL Processes and Transformation between Different Data Models}

Since ETL processes are a widely used technology in the database field, developed to load data from various data sources
into a data warehouse or relational databases, numerous methods are already available. The most challenging aspect of ETL is the \textit{transformation} step. In our approach, we aim to transform data between two different database models (the input data model and the relational database as the target database) and simultaneously to transform between the structures of the source and target data.

For relational databases, 
ETL scripts have to be developed for each new input data set. 
%It is therefore not surprising that 
Numerous solutions are available in this area, ranging from specialized ETL tools like Qlik Talend\footnote{www.talend.com} to components in systems such as DataBricks\footnote{www.databricks.com/}, and data integration tools like Altova MapForce\footnote{www.altova.com/mapforce}.
They all provide a graphical user interface to speed up the matching process as well as libraries containing certain mapping functions. Even though this makes programming ETL scripts faster, programming skills are still required.

\paragraph{Transformations between different data models}
An integral part of these ETL tools is the \textit{transformation between different data models}. These methods have always been developed when new data models came up. So there have been approaches for translations between relational databases and XML  \cite{tatarinov2002storing,amer2004comprehensive,carey2000xperanto,
%KlettkeMeyerWebdb02,
KlettkeMeyer2003} and between
relational databases and JSON \cite{DBLP:conf/webdb/ChasseurLP13,liu2014json,DBLP:conf/ACMse/BahtaA19,DBLP:conf/iiwas/Petkovic20,DBLP:journals/tkde/ChillonKRM24}. These transformation tasks are also integral part of each multi-model approach \cite{lu2019multi,DBLP:conf/vldb/HolubovaKS19, DBLP:conf/models/ConradUKS22,DBLP:conf/er/HauslerCSSK25} and are available in many data integration tools like Qlik\footnote{www.qlik.com}, MapForce\footnotemark[4] and AWS Glue\footnote{aws.amazon.com}.

%The easier transformation direction is from structured data models into semi-structured data models. 
%For the transformation from semistructured to structured data models, there exist several methods -- usually a structured approach \cite{DBLP:conf/ACMse/BahtaA19}, as datatype in the relational model \cite{liu2014json}, or a node-wise approach \cite{tatarinov2002storing,DBLP:conf/iiwas/Petkovic20}. 
%Also OR mappers like Hibernate\footnote{hibernate.org} and SQLAlchemy\footnote{www.sqlalchemy.org/}  offer different variants on how to transform data, see \cite{TORRES20171} for an overview. 

We have seen for several years now that data is not only provided in relational formats; the discussions surrounding big data and NoSQL databases 
%, which began more than a decade ago, 
underscore this point. Recent work such as \cite{DBLP:conf/edbt/0001VPN24} also shows the need to consider data from various formats as input data. For these reasons, we consider various data formats as input data, such as relational data, CSV, JSON, and graph data. Accordingly, the above mentioned transformation into JSON format can be used and is integrated in NovaFilter.

\paragraph{Schema matching and mapping}
Parallel, there exist techniques for \textit{schema matching and mapping}. They require a source and target schema as input and use distance functions to find a matching, from which in a second step a mapping  is determined to transform the data from one schema to another.
Numerous methods are known, such as \cite{christen2012data} for relational data, \cite{lee2002xclust} for hierarchical XML data, \cite{khatiwada2022integrating} for datasets with different data models in a data lake, and the well-known overview of different classes of methods \cite{rahm2004matching}. 
In the context of evolution, we have also developed a
matching and mapping approach for JSON \cite{DBLP:conf/bigdataconf/KlettkeAS0S17}. Furthermore, currently there are several other approaches under development that outsource the transformation part to an AI component that's why, we do not enumerate the single publications here but want to mention the field.

\paragraph{Rule-based Constraint Checking and Data Correction in Semi-structured Data} 

Two of the central parts of the project are a constraint check and a rules-based transformation approach. Here, we can base on solutions from other computer science fields. 
The \textit{separation of logic and program code}
to ensure flexibility and adaptability is one of the fundamental concepts in computer science. It is applied for instance in (i) model-driven approaches, to define rules for translating between an abstract and a system-specific model \cite{mens2006taxonomy,czarnecki2003classification}, (ii) transformations between different data models,  e.g. between XML and relational databases \cite{carey2000xperanto,amer2004comprehensive}, (iii) between different layers of a software architecture, e.g. XSLT as a prominent example 
for a separate specification of layout transformations 
\cite{clark1999xsl}, and (iv) data cleaning %or correction 
rules %in data engineering 
\cite{boeckling2022cleaning}, often through (partial) functional dependencies \cite{ChuIlyas2013, DBLP:books/acm/IlyasC19}.  
In all cases, rules are defined with left-hand side (the matching part) and right-hand side, the correction or transformation part. 

%In all cases, rules are defined with left-hand side (the matching part) and right-hand side, the correction or transformation part. The left-hand side specifies, in which cases, the rule will be applied, the right-hand side contains either the new information for the correction or transformation. 

The remaining research gap that NovaFilter will fill is to automate the validation of relational structure and integrity constraints in the staging area and to combine a rule-based method with an interactive approach and to examine their application in ETL contexts.

%In this project, we have to prove feasibility of a rule-based approach for the ETL context and to develop rules for constraint checking  and for transformations (e.g. standard methods like type cast and user-defined rules for mappings). 

\paragraph{Human-in-the-loop Approaches in Data Engineering}

NovaFilter will integrate two different human-in-the-loop approaches: 
(i) the example discussion with users from which new transformation rules will be derived and (ii) the \textit{explainability components}, the generation of a log file and an interactive approach allowing users to monitor the transformations.

In data engineering, human-in-the-loop approaches are used for database modeling~\cite{Gomez2026}, schema inference \cite{baazizi2020human,DBLP:conf/esws/SadruddinDPWGRKAMK25}, JSON and graph data exploration \cite{DBLP:conf/btw/MollerBKSS19,DBLP:conf/models/HauslerK24}, data integration \cite{li2017human}, monitoring of data pipelines \cite{grafberger2021mlinspect,shankar2022, DBLP:conf/hilda/HasanP25,DBLP:conf/hilda/StrasserK24,DBLP:conf/btw/LandesKK25}, and many other scenarios. 
%In Data Engineering Processes, human-in-the-loop approaches are frequently used. 
Not only the numerous publications at conferences demonstrate how widespread this concept is, the workshop series HILDA\footnote{HILDA@SIGMOD -- Workshop on Human-In-the-Loop Data Analytics} and DaSH\footnote{DaSH@VLDB -- Data Science with Human-in-the-Loop} also underline this.

%At the first glance, the planned human-in-the-loop approaches do sound like implementation tasks in the software architecture but we expect several challenges that have to be solved in the project.
The available approaches will inspire the dialog design for exam\-ple-based program transformation. 
Of particular importance here are (i) the development of a clear transformation-by-example approach,
(ii) the dashboard development for monitoring the overall processes, and (iii) an interactive explanation component that offers various levels of abstraction.

\section{Conclusion}

Yes, I know NovaFilter is still in an early stage of development   and I’d like to share its idea with the database community. I think NovaFilter is an interesting alternative to the current mainstream LLM-based approaches. It is exactly the opposite: a fully rule-based approach that uses example-based discussions with users to continuously expand the rule set.

This enables a domain-specific definition of the mapping rules, and developers don’t have to write ETL scripts—they simply specify the desired mappings by discussing examples. NovaFilter is therefore a low-code application.
In my opinion, this approach 
has the potential for a completely new method class in ETL script generation.

The early stage is also evident in the fact that while the implementation of the first two stages (rule-based transformation) is in place, the third stage (user interaction) is not yet available. And already, these first two stages solve the same tasks that ETL scripts perform. Compared to ETL scripts, they even offer the added value of a separately externalized set of rules that describes this mapping. 

With these first two stages, it is therefore possible to map various data formats to relational databases, perform transformations for simple data types, and carry out mapping using predefined rules. 
This means that all transformations affecting the schema can already be implemented.
%I would like to further develop this idea in the future and create a tool that is formally precise, requires little programming knowledge, and produces results that are justifiable and reproducible. 

One can argue that NovaFilter only shifts the effort: from manually programming ETL scripts to discussion examples. This is true but this shift has a lot of advantages: programming ETL scripts require programming skills as well as knowledge about the domain that's why it requires often an interdisciplinary cooperation between a programmer and a domain expert. 
For correcting examples, one only needs to have and to use their domain knowledge. So, this shift enables that domain experts can do with a tool support realise the ETL definitions.   

\section{Future Work}
The next steps are very clearly defined in this project. The first task is to continue \textit{implementing Stage 3}: interacting with users to derive new rules from example transformations. This will be followed by an evaluation based on various scenarios, different data models, and various available data from benchmarks. 
From this, we will determine the system’s various \textit{learning curves} for \textit{different heterogeneity classes}. 
An \textit{empirical study} with users from various fields—such as biology, medicine, economics, and history
will follow to optimize the dialogue generation and the user interface.  

Currently, we start for each new target database with an empty set of user-defined transformation rules and extend the rule base in the process. 
Another idea is to collect all transformation rules in a database that serves as a \textit{knowledge base} for the approach. A \textit{recommendation system} built on this knowledge base shall then be used to generate transformation rules for new target databases. To guarantee data protection, it has to be ensured that only data from the same person or company is used for the recommendations. With this approach, we aim to \textit{mitigate the existing cold-start problem}.  

The final task for now regarding future work is to extend the rule set to include conditional rules. In this process, conditions must be taken into account in the rules. These are derived from the context of the examples entered by the user. The example shown in this article prepared the ground for this approach. A conditional transformation rule would be the following:   

\noindent
"aliases": 
\{"price": [if "pricing/currency" = "EUR":
"pricing/amount"]\}.

\section*{Acknowledgment}
I would like to thank the student groups 
working on the software projects 
who were infected by the idea and took the first steps toward implementation in creative ways. Special thanks go to Tom Schönleben and Korbinian   
Staudacher, the scrum masters of the teams.

\bibliographystyle{plain}
\bibliography{bibliography,bib_klettke}  
\end{document}